\documentclass[fleqn,twoside]{article}
\usepackage{espcrc2}

\usepackage{graphicx}
\usepackage{epsfig}
\usepackage[figuresright]{rotating}

\newcommand{\eqn}[1]{Eqn.~\ref{#1}}
\newcommand{\tab}[1]{Table~\ref{#1}}
\newcommand{\fig}[1]{figure~\ref{#1}}

\newcommand{\AmS}{{\protect\the\textfont2
  A\kern-.1667em\lower.5ex\hbox{M}\kern-.125emS}}

\title{Baryon operators and spectroscopy in lattice QCD}

\author{LHP Collaboration:
	S.~Basak\address[umd]{Department of Physics, 
        University of Maryland, College Park, MD 20742, USA},
	R.~Edwards\address[jlab]{Thomas Jefferson National Accelerator 
	Facility, Newport News, VA 23606, USA},
	R.~Fiebig\address{Physics Department, Florida International University,
	Miami, FL 33199, USA},
	G.~Fleming\addressmark[jlab],
	U.M.~Heller\address{American Physical Society, One Research Road,
	Ridge, NY 11961-9000, USA},
	C.~Morningstar\address{Department of Physics, Carnegie Mellon
	University, Pittsburgh, PA 15213, USA},
	D.~Richards\addressmark[jlab],
	I.~Sato\addressmark[umd]and
	S.~Wallace\addressmark[umd].}
       
\begin{document}

\begin{abstract}
The construction of the operators and correlators required to
determine the excited baryon spectrum is presented, with the aim of
exploring the spatial and spin structure of the states while
minimizing the number of propagator inversions.  The method used to
construct operators that transform irreducibly under the symmetries of
the lattice is detailed, and the properties of example operators are
studied using domain-wall fermion valence propagators computed on MILC
asqtad dynamical lattices.
\vspace{1pc}
\end{abstract}
\maketitle

\section{INTRODUCTION}

The determination of the excited baryon spectrum is an important step
on the road to understanding the dynamics of QCD, and it is, for example,
a vital element of the Jefferson Laboratory experimental program.
Lattice gauge theory has a pivotal role not only in providing
\textit{ab initio} calculations of the masses of the lowest lying
states, but also in establishing the quark and gluon content of the
states.  There has thus been increasing activity amongst the lattice
community that has demonstrated that the masses of the lowest-lying
baryon states of both parities for spin-$1/2$ and spin-$3/2$ are
accessible to lattice calculation, and there have been calculations of
the mass of the first positive-parity, spin-$1/2$ excitation of the
nucleon, the so-called Roper resonance.

Most of these calculations have been made in the quenched approximation to
QCD, and an important component of future lattice studies will be a
careful analysis of the systematic uncertainties on these results.
However, the confrontation of lattice calculations with experiment
will require a more complete understanding of the spectrum, obtaining
both the masses of the higher spins and of the radial excitations.
Furthermore, the discovery of the quark and gluon structure of the
excited states will require a more extensive basis of interpolating
operators than used so far, allowing both for excited glue and for
multi-quark operators.  The aim of this talk is to describe the design
of such operators, and in particular to construct operators that
transform irreducibly under the symmetries of the lattice.

The remainder of this talk is laid out as follows.  In the next
section, the methodology of computing the spectrum is described, and
the symmetries used to classify baryon states are detailed.
Gauge-invariant ``elemental'' operators that have the correct flavor
structure and that explore the spatial structure of the baryons are
then introduced.  The reduction of these operators to ones that
transform irreducibly under the cubic symmetry of the lattice is then
performed.  Finally, the reduction is illustrated for the case of
point-like elemental operators obtained using domain-wall fermion
(DWF) valence propagators computed on a dynamical asqtad-fermion
background.

\section{THE SPECTRUM FROM LATTICE CALCULATIONS}

The computation of the spectrum of states in lattice QCD is in
principle straightforward.
\begin{enumerate}
\item Choose an interpolating operator ${\cal O}$ that has a good
overlap with $P$, the state of interest,
\[
\langle 0\mid {\cal O} \mid P\rangle \ne 0,
\]
and ideally a small overlap with other states having the same quantum
numbers.
\item Form the time-sliced correlation function
\[
C(t) = \sum_{\vec{x}} \langle {\cal O}(\vec{x}, t) {\cal O}^{\dagger}
(\vec{0}, 0) \rangle .\label{eq:corrs_cons}
\]
\item  Insert a complete set of states between ${\cal O}$ and ${\cal
O}^{\dagger}$.  The time-sliced
sum puts the intermediate states at definite momentum, and we find
\begin{eqnarray*}
\lefteqn{C(t) =
 \sum_{\vec{x}} \sum_P \int \frac{d^3k}{(2 \pi)^3 2 E(\vec{k})}}\\
& & \times \langle 0 |
{\cal O}(\vec{x}, t) | P(\vec{k}) \rangle \langle P(\vec{k}) | {\cal
O}^{\dagger} (\vec{0},0) | 0 \rangle\\
 &  & = \sum_P \frac{\mid \langle 0 \mid {\cal O} \mid P
\rangle \mid^2}{2 \, E_P(\vec{0}) } e^{i E_P(\vec{0}) t}.
\end{eqnarray*}
where the sum over $P$ includes the contributions from two-particle
and higher states.
\item Continue to Euclidean space $t \rightarrow i t$, yielding
\begin{equation}
C(t) = \sum_P \frac{\mid \langle 0 \mid {\cal O} \mid P
\rangle \mid^2}{2 \, m_P} e^{- m_P t},\label{eq:spectral_sum}
\end{equation}
so that the correlator falls off exponentially with the mass of the
lightest state at large times.
\end{enumerate}

In order to obtain the mass of the higher resonances in
\eqn{eq:spectral_sum}, the most appealing approach is to apply
variational methods\cite{michael85,lw90}, beginning with the
computation of a matrix of correlators
\begin{equation}
 C_{ij}(t) = \sum_{\vec{x}} \langle {\cal O}_i(\vec{x},t) {\cal
 O}^{\dagger}_j(0) \rangle \label{eq:matrix_corr},
\end{equation}
where the $\{{\cal O}_i\}$ form a basis of operators of definite
quantum numbers.  In the usual application of the variational
technique, we find a basis of eigenvectors $V(t_0)$ that diagonalizes the
transfer matrix $C(t_0)^{-1} C(t)$ for $t = t_0 + 1$ for some $t_0$
close to the source, and determine the spectrum from the eigenvalues of
\begin{equation}
V(t_0+1)^{-1} C(t_0)^{-1} C(t) V(t_0+1)\label{eq:variation}
\end{equation}
at subsequent times.  The method relies on constructing a basis of
operators $\{{\cal O}_i\}$ that provides a good description of the
states of interest, and the remainder of this talk will focus on the
construction of that basis.

We classify the states of the continuum by their flavor structure,
$F$, and by their parity $P$, total spin $J$ and ``helicity'' $J_z$.
Whilst the flavor structure can be faithfully preserved in a lattice
calculation, the parity, spin and helicity labels arise from looking at the
symmetry properties of states under rotations and reflections.  In a lattice
calculation, we have replaced continuum space time by a hypercubic
lattice, and thus must classify states according to the irreducible
representations $\Lambda$ of the cubic group (for particles at rest);
the row $\lambda$ within the representation is then the analogue of helicity.

In the following section, we will construct a set of elemental baryon
operators $\{B^F_i(x)\}$ having the correct flavor properties $F$. We
will then use group theory to generate a set of baryon operators $\{
B^{F\Lambda\lambda}_i(x)\}$ that transform irreducibly under the
rotation and reflection symmetries of the lattice, and from which we
will construct our correlation matrix
\begin{equation}
C^{\Lambda\lambda F}_{ij}(t) = \sum_{\vec{x}} \langle
B^{\Lambda\lambda F}_i(x) \overline{B}^{\Lambda\lambda F}_j(0)
\rangle,\label{eq:corr_mat}
\end{equation}
and hence extract the spectrum of states corresponding to $\Lambda, \lambda$.

\section{ELEMENTAL BARYON OPERATORS}
The color and flavor structure of the operators is dictated by the
requirements of gauge invariance and of isospin respectively; it is
natural in a lattice calculation to assume exact isospin symmetry, and
to classify states and operators by their isospin and strangeness,
rather than according to $SU(3)$ flavor.

The use of ``smearing'' has proven essential to reduce the coupling of
operators to the higher excited states, and thus to ensure that the
correlators \eqn{eq:matrix_corr} are dominated by only a few states
even close to the source.  In the example given here, we will adopt a
gauge-covariant spatial Laplacian to smear the fields:
\begin{equation}
\tilde{\psi}(x) = (1 + \sigma^2 \tilde{\Delta}/4N )^N \psi(x)\label{eq:smear}
\end{equation}
with the three-dimensional Laplacian defined by
\begin{equation}
   \tilde{\Delta} \psi(x) = \sum_{k=\pm 1, \pm 2,\pm 3} \left(
   \tilde{U}_k(x)\psi(x\!+\!\hat{k})-\psi(x) \right),
\end{equation}
where $\tilde{U}$ denotes a link variable that can be smeared
according to, say, the APE prescription\cite{APEsmear}.  Both $N$ and
$\sigma$ are tunable parameters, and in the limit $N \longrightarrow
\infty$, \eqn{eq:smear} reduces to Gaussian smearing of width $\sigma$.
Note that the square of the smeared field vanishes, in the manner of a
simple Grassmann field.

An important consideration is to construct operators that enable us to
build up the radial and orbital structure of the states.  This is
accomplished by additional powers of the Laplacian, and by spatial
derivatives, respectively, so that the quark building blocks become:
\begin{eqnarray}
\chi^n_{Aa\alpha}(x)&\equiv& 
  \left( \tilde{\Delta}^{n} \tilde{\psi}(x)\right)_{Aa\alpha},\\
\xi^{npj}_{Aa\alpha}(x)&\equiv& \left(\tilde{D}^{(p)}_j\!\tilde{\Delta}^{n} 
           \tilde{\psi}(x)\right)_{Aa\alpha},
\end{eqnarray}
where $A$ is a flavor index, $a$ is a color index, $\alpha$ a Dirac
spin index, and $n$ is an integer.  The $p$-link gauge-covariant
forward displacement operator is defined by
\begin{eqnarray}
\lefteqn{\tilde{D}_j^{(p)}O(x) = }\nonumber\\
& &  \tilde{U}_j(x)\dots 
   \tilde{U}_j(x\!+\!(p\!-\!1)\hat{j}) O(x\!+\!p\hat{j}),
\end{eqnarray}
where $j = 1,2,3$, and $p$ is an integer specifying the length of the
displacement.

From these quark building blocks, we construct our elemental
three-quark operators:
\begin{eqnarray}
 & & \phi^F_{ABC}\ \varepsilon_{abc} 
   \ \chi^{n_1}_{Aa\alpha} 
   \ \chi^{n_2}_{Bb\beta}
   \ \chi^{n_3}_{Cc\gamma}, \\
  & & \phi^F_{ABC}\ \varepsilon_{abc}
 \ \chi^{n_1}_{Aa\alpha}
 \ \chi^{n_2}_{Bb\beta}
 \ \xi^{n_3 pj}_{Cc\gamma}, \\
  & & \phi^F_{ABC}\ \varepsilon_{abc}
 \ \chi^{n_1}_{Aa\alpha} 
 \ \xi^{n_2 p_1 j}_{Bb\beta}
 \ \xi^{n_3 p_2 k}_{Cc\gamma}.
\end{eqnarray}
The flavor structure is specified by $\phi^F_{ABC}$, whilst the
Levi-Civita symbol ensures that these are color singlets.  The radial
and orbital structures are explored by varying $n_1, n_2, n_3$, and
through the choices of $p_1, p_2, j, k$ respectively. A large number
of baryons can then be studied using a somewhat small number of quark
propagator sources.

\section{SYMMETRIES OF THE LATTICE}

Proper rotations restricted to an isotropic cubic lattice form the
cubic group $O$. It has 24 elements, and five conjugacy classes
and thus five single-valued irreducible representations: $A_1, A_2, E,
T_1, T_2$, of dimensions $1, 1, 2, 3~\mbox{and}~3$ respectively.  If
we allow for spatial inversions, corresponding to $P = \pm 1$, we
obtain the group $O_h$, and the irreducible representations acquire a
further label, $g$ or $u$, corresponding to positive and negative
parity states, respectively.  The construction of operators
transforming irreducibly according to the single-valued
representations is well known, and has been crucial to identifying the
states in lattice calculations of the glueball spectrum.

There are four two-dimensional spinorial representations of $O_h$:
$G_{1g}, G_{1u}, G_{2g}~\mbox{and}~G_{2u}$, and two four-dimensional
representatons, $H_g~\mbox{and}~H_u$\cite{johnson,mandula2}. The
irreducible representations $J$ of the continuum group $SU(2)$ are
reducible under the cubic group $O$; the number of times
$n^J_{\Gamma}$ that each of these reducible representations occurs in
the irreducible representation $\Gamma$ of $O$ is shown in
\tab{tab:subduction}; the extension to $O_h$ is straightforward.
\begin{table}[htb]
\caption{The number of times $n^J_{\Gamma}$ that the irreducible
  representation $\Gamma$ of $O$ occurs in the reduction of the
  irreducible representation $J$ of $SU(2)$.}
\label{tab:subduction}
\renewcommand{\tabcolsep}{1.5pc} 
\renewcommand{\arraystretch}{1.2} 
\begin{tabular}{rrrr}
\hline
$J$ & $n^J_{G_1}$ & $n^J_{G_2}$ & $n^J_{H}$\\
\hline
$1/2$ & 1 & 0 & 0\\
$3/2$ & 0 & 0 & 1\\
$5/2$ & 0 & 1 & 1\\
$7/2$ & 1 & 1 & 1\\
$9/2$ & 1 & 0 & 2\\
\hline
\end{tabular}
\end{table}

Whilst full rotational symmetry is restored in the continuum limit,
the irreducible representations of $O_h$ contain states of many
representations of $SU(2)$, and in general the degrees of
freedom, corresponding to different helicities, lie in different
irreducible representations.  Thus, for example, a state of spin $5/2$
has four degrees of freedom in the four-dimensional representation
$H$, and two degrees of freedom in the two-dimensional representation
$G_2$; the identification of the spin of a state is accomplished by
looking for the approach to a common mass across the different
irreducible representations in the continuum limit.

Central to the task of finding the $B^{\Lambda\lambda F}_i(x)$ is the
projection formula
\begin{equation}
  B_i^{\Lambda\lambda F}\!(\vec{x})\!
 = \!\frac{d_\Lambda}{g_{O_h}}\!\!\!\sum_{R\in O_h}\!\!
  \!\!D^{(\Lambda)\ast}_{\lambda\lambda}(R)
   U_R B^F_i\!(\vec{x}) U_R^\dagger,
\label{eq:project}\end{equation}
where $\Lambda$ refers to an $O_h$ irrep, $\lambda$ is the irrep row,
$g_{O_h}$ is the number of elements in $O_h$,
$d_\Lambda$ is the dimension of the $\Lambda$ irrep,
$D^{(\Lambda)}_{mn}(R)$ is a $\Lambda$
representation matrix corresponding to group element $R$,
and $U_R$ is the quantum operator which implements the symmetry
operations; the temporal argument is suppressed.

Application of this formula requires explicit representation matrices
for every group element.  Representation matrices for all allowed
proper rotations can be generated from the matrices for $C_{4y}$ and
$C_{4z}$, the rotations by $\pi/2$ about the $y$- and $z$-axes,
respectively.  The explicit matrices we use for these generators are
given by
\[
 D^{(G_1)}(C_{4y})=\frac{1}{\sqrt{2}}\!\left[\begin{array}{rr}
 \!\!1 & \!\!\!\!-1\!\! \\ \!\!1 & \!\!\!\! 1\!\!
\end{array}\right]=-D^{(G_2)}(C_{4y}),
\]
\[
 D^{(G_1)}(C_{4z})=\frac{1}{\sqrt{2}}\!\left[\begin{array}{cc}
  1\!-\!i & \!\!\!\!0 \\ 0 &\!\!\!\! 1\!+\!i 
\end{array}\right]\!\! = -D^{(G_2)}(C_{4z}),
\]
\[
 D^{(H)}(C_{4y})=\frac{1}{2\sqrt{2}}\left[\begin{array}{rrrr}
  \!1 \!&\! -\sqrt{3} \!&\! \sqrt{3} & -1 \!\\
  \!\sqrt{3} \!&\! -1 \!&\! -1 \!&\!  \sqrt{3} \!\\
  \!\sqrt{3} \!&\!  1 \!&\! -1 \!&\! -\sqrt{3}\! \\
  \!1 \!&\!  \sqrt{3} \!&\! \sqrt{3} &  1\! \end{array}\right],
\]
\[
 D^{(H)}(C_{4z})=\frac{1}{\sqrt{2}}\left[\begin{array}{cccc}
   \!\!-1\!-\!i \!&\! 0 \!&\! 0 & 0 \!\!\\
   \!\!0 \!&\! 1\!-\!i \!&\! 0 & 0 \!\!\\
   \!\!0 \!&\! 0 \!&\! 1\!+\!i & 0 \!\!\\
   \!\!0 \!&\! 0 \!&\! 0 \!&\! -1\!+\!i \!\!\end{array}\right].
\]

\section{EXAMPLE: LOCAL NUCLEON INTERPOLATING OPERATOR}

We will illustrate the procedure above by examining the reduction of
the local proton interpolating operator, with $I=1/2, I_3 = 1/2$.
The elemental operators that are gauge invariant and have the correct
isospin are
\begin{equation}
\Phi_{\alpha\beta\gamma} = \epsilon^{abc}(u^a_{\alpha} d^b_{\gamma}
u^c_{\beta} - d^a_{\alpha} u^b_{\gamma} u^c_{\beta}).\label{eq:phi}
\end{equation}
Examination of \eqn{eq:phi} reveals the constraints
$\Phi_{\alpha\beta\gamma} + \Phi_{\gamma\beta\alpha} = 0$ and
$\Phi_{\alpha\beta\gamma} + \Phi_{\beta\gamma\alpha} +
\Phi_{\gamma\alpha\beta} = 0$ so that there are only 20 independent
operators.  The projection operation \eqn{eq:project} is applied to
each of the twenty linearly independent operators in \eqn{eq:phi}, and
linearly independent operators identified that transform
irreducibly under $O_h$.  The procedure depends on the Dirac basis
adopted, and therefore the reduction must be performed using the basis
employed to compute the quark propagators.  The reduction in the
DeGrand-Rossi basis is shown in \tab{tab:results}; the corresponding
reduction in the Dirac basis is provided in ref.~\cite{morningstar03}.
\begin{table*}[t]
\caption{Combinations of the operators $\Phi_{\alpha\beta\gamma}$
 in \protect\eqn{eq:project} 
 which transform irreducibly under $O_h$ for the DeGrand-Rossi
 representation of the $\gamma$-matrices, employed by LHPC and MILC.
\label{tab:results}}
\begin{center}
\renewcommand{\tabcolsep}{4mm} 
\begin{tabular}{ccc}\hline
 Irrep & Row &  Operators
    \\ \hline 
  $G_{1g}$  & 1 &  $\Phi_{112} + \Phi_{334}$  \\
  $G_{1g}$  & 2 &  $-\Phi_{221} - \Phi_{443}$  \\ \hline
  $G_{1g}$  & 1 &  $\Phi_{123} - \Phi_{213} + \Phi_{314}$\\
  $G_{1g}$  & 2 &  $\Phi_{124} - \Phi_{214} + \Phi_{324}$  \\ \hline
  $G_{1g}$  & 1 &  $2\Phi_{114} + 2\Phi_{332} - \Phi_{123} -
    \Phi_{213} + 2\Phi_{134} - \Phi_{314}$ \\ 
  $G_{1g}$  & 2 &  $-2\Phi_{223} - 2\Phi_{441} + \Phi_{124} -
    \Phi_{214} - 2\Phi_{234} + \Phi_{324}$  \\ \hline
  $G_{1u}$  & 1 &  $\Phi_{112} - \Phi_{334}$  \\ 
  $G_{1u}$  & 2 &  $-\Phi_{221} + \Phi_{443}$  \\ \hline
  $G_{1u}$  & 1 &  $\Phi_{123} - \Phi_{213} - \Phi_{314}$  \\
  $G_{1u}$  & 2 &  $\Phi_{124} - \Phi_{214} - \Phi_{324}$  \\ \hline
  $G_{1u}$  & 1 &  $2\Phi_{114} - 2\Phi_{332} - \Phi_{123} -
    \Phi_{213} - 2\Phi_{134} + \Phi_{314}$  \\ 
  $G_{1u}$  & 2 &  $-2\Phi_{223} + 2\Phi_{441} + \Phi_{124} +
    \Phi_{214} + 2\Phi_{234} - \Phi_{324}$  \\ \hline
  $H_{g}$   & 1 &  $\sqrt{3}(\Phi_{113} + \Phi_{331})$\\
  $H_{g}$   & 2 &  $\Phi_{114} + \Phi_{332} + \Phi_{123} +
    \Phi_{213} - 2\Phi_{134} + \Phi_{314}$  \\
  $H_{g}$   & 3 &  $\Phi_{223} - \Phi_{441} + \Phi_{124} +
    \Phi_{214} - 2\Phi_{234} + \Phi_{324}$  \\
  $H_{g}$   & 4 &  $\sqrt{3}(\Phi_{224} + \Phi_{442})$ \\ \hline
  $H_{u}$   & 1 &  $\sqrt{3}(\Phi_{113} - \Phi_{331})$\\ 
  $H_{u}$   & 2 &  $\Phi_{114} - \Phi_{332} + \Phi_{123} +
    \Phi_{213} + 2\Phi_{134} - \Phi_{314}$ \\
  $H_{u}$   & 3 &  $\Phi_{223} - \Phi_{441} + \Phi_{124} +
    \Phi_{214} + 2\Phi_{234} - \Phi_{324}$  \\
  $H_{u}$   & 4 &  $\sqrt{3}(\Phi_{224} - \Phi_{442})$ \\ \hline
\end{tabular}
\end{center}
\end{table*}

There are three embeddings of $G_1$ and a single embedding of $H$, of
both parities.  This is in accord with expectations from an analysis
of the continuum operators, where it is possible to construct three
linearly independent nucleon interpolating currents\cite{brommel03},
and a single spin-$3/2$ interpolating current.

The computation of the correlator matrix \eqn{eq:corr_mat} is
straightforward. We compute the
gauge-invariant ``generalised'' baryon correlator
\begin{eqnarray}
\lefteqn{B_{\alpha\beta\gamma,\bar{\alpha}\bar{\beta}\bar{\gamma}}(t;0)
=}\nonumber\\ & & \sum_{\vec{x}}
\epsilon^{abc}\epsilon^{\bar{a}\bar{b}\bar{c}}
Q_{\alpha\bar{\alpha}}^{a\bar{a}} (x; 0)
Q_{\beta\bar{\beta}}^{b\bar{b}} (x; 0)
Q_{\gamma\bar{\gamma}}^{c\bar{c}} (x; 0)
\end{eqnarray}
where $Q$ is the quark propagator, whence
\begin{equation}
\sum_{\vec{x}} \langle \Phi_{\alpha\beta\gamma}(\vec{x},t)
\bar{\Phi}_{\bar{\alpha}\bar{\beta}\bar{\gamma}} \rangle =
\tilde{B}_{\alpha\beta\gamma,\bar{\alpha}\bar{\beta}\bar{\gamma}},
\end{equation}
where $\tilde{B}$ is obtained from $B$ by symmetrizing over $\alpha$
and $\beta$ to account for the two possible Wick contractions, and
antisymmetrizing over $\alpha,\gamma$ and over $\bar{\alpha},
\bar{\gamma}$ to take account of the two terms in \eqn{eq:phi}.
Finally, we form the correlators
\begin{equation}
C_{ij}(t) = \sum_{\vec{x}}\Gamma_i^{\alpha\beta\gamma}
\bar{\Gamma}_j^{\bar{\alpha}\bar{\beta}\bar{\gamma}} \langle \Phi_{\alpha\beta\gamma}(\vec{x},t) 
\bar{\Phi}_{\bar{\alpha}\bar{\beta}\bar{\gamma}}(0)\rangle,\label{eq:corrs}
\end{equation}
where the $\Gamma_i,\bar{\Gamma}_j$ are given in \tab{tab:results}.
While the computation of the generalized baryon correlators is
computationally quite demanding, the subsequent calculation of the $20
\times 20$ matrix of correlators of \eqn{eq:corrs} can be performed in
a matter of minutes on a workstation, even for several hundred
configurations.

It is instructive to list some of the properties of the correlation
matrix \eqn{eq:corrs}:
\begin{itemize}
\item Cross-correlations between operators within different
  irreducible representations vanish.
\item Cross-correlations between the operators of different rows
  within the same representation vanish.
\item The different rows within a representation can be chosen to have
  the same normalisation, so that correlators thus constructed have
  the same expectation values.
\item Cross-correlations between the same row of different embeddings
  of the same irreducible representation \textit{do not vanish};
  they can be constructed to vanish only on a particular time slice.
\end{itemize}
Thus the strategy for extracting the mass spectrum from the data is
now clear.  Firstly, choose a particular irreducible representation,
possibly averaging over the rows within that representation.  Then form the
matrix of correlators between all the averaged operators within that
representation, and extract the spectrum using, say, the method of
\eqn{eq:variation}.

The talk concludes with an examination of the correlators on an
ensemble of lattice data, and in particular a verification of the
properties above.  To perform this study, we employ an ensemble of
around 300 configurations computed on a $20^3 \times 64$ lattice with
$N_f = 2 + 1$ flavors of asqtad staggered quarks having $a m_u/ a m_s =
0.01/0.05$\cite{milc99}.  The lattice spacing $a = 0.124~{\rm fm}$ is
obtained from the 1P-1S splitting in the Upsilon
spectrum\protect\cite{gray2002}.

We employ domain-wall fermions at a single mass $m = 0.01$ for the
valence quarks, which are computed on ``chopped'' $20^3 \times 32$
lattices, with $N_5 = 16$; to reduce the residual mass, the
configurations are first blocked using a non-perturbative HYP blocking
scheme\cite{hyp}.  Smeared sources are used, using $\sigma = 4.35$ and $N = 30$
in \eqn{eq:smear}, but local sinks, and thus the correlators are not
positive definite; full details of the calculation will appear in a
subsequent paper~\cite{lhpc_milc}.

\begin{figure}[htb]
\psfig{width=7cm,file=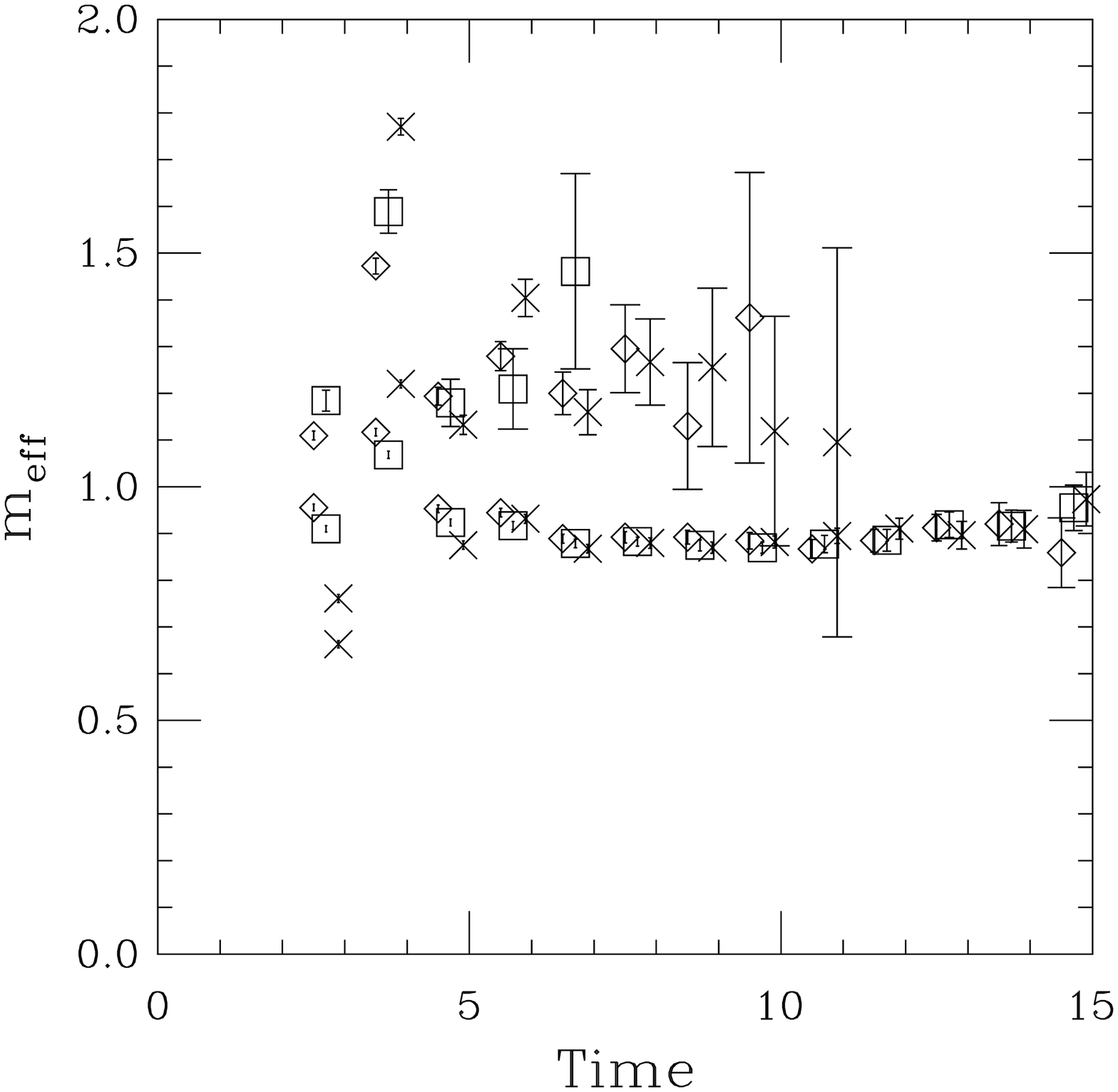}\\
\psfig{width=7cm,file=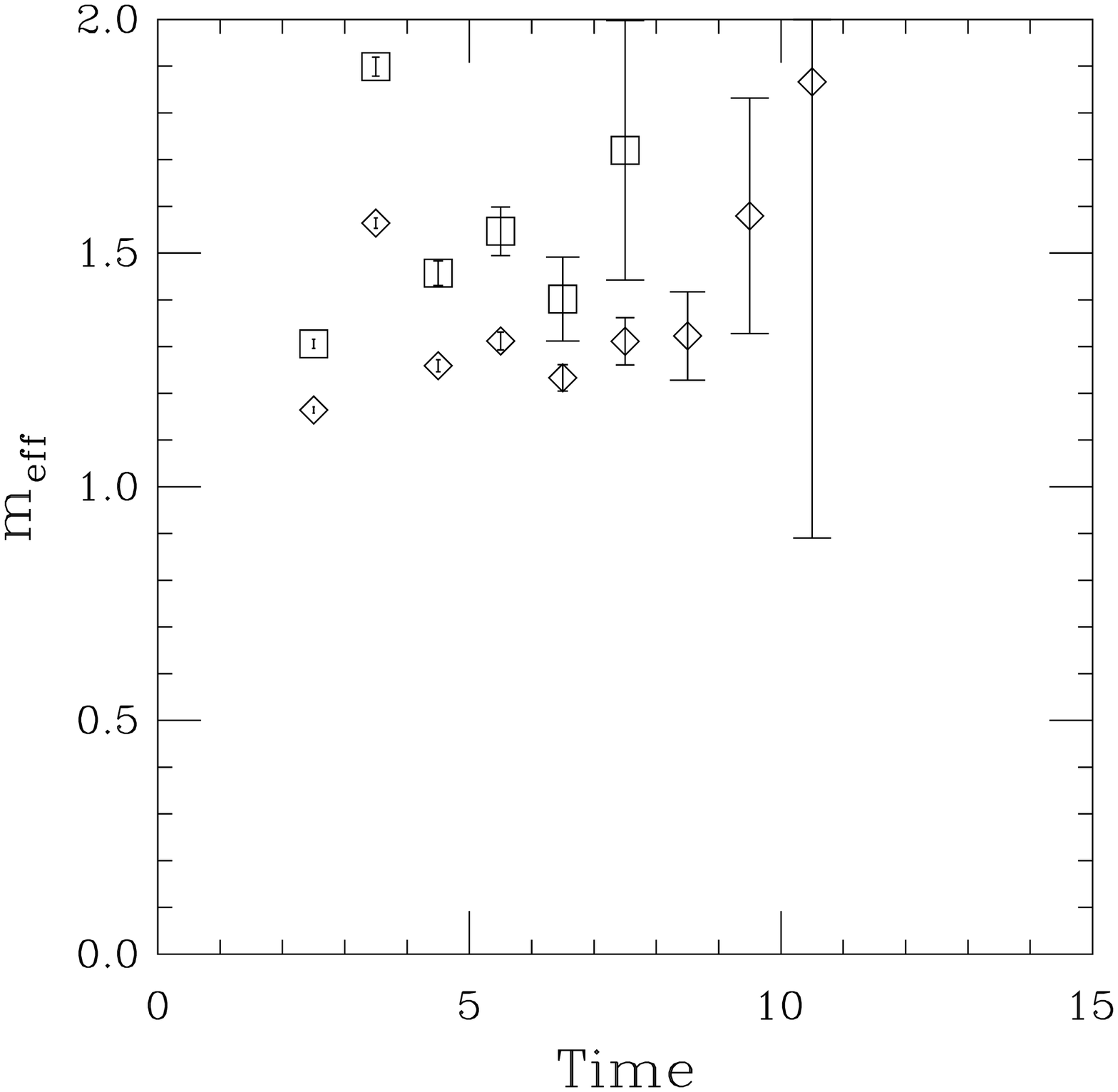}
\caption{The upper figure shows the effective masses for the three
  different embeddings of $G_{1g}$ (smaller) and $G_{1u}$ (larger)
  as the three different plotting symbols, and valence quark mass $m =
  0.01$.  The points corresponding to the different embeddings are
  offset for clarity.  The lower figure shows the effective masses for
  $H_u$ (diamond) and $H_g$ (square).}
\label{fig:g1_mass}
\end{figure}
The effective masses obtained from the diagonal correlators using the
three embeddings of $G_{1g}$ and $G_{1u}$ are shown in
\fig{fig:g1_mass}.  The plot reveals a clear signal for both positive-
and negative parity states from each of the three embeddings.  Note
that the three usual baryon interpolating operators, for example in
ref.~\cite{brommel03}, correspond to linear combinations of the
operators used here; the identification of a ``good'' baryon operator
corresponding to the nucleon, and a baryon operator having a good
overlap onto the Roper resonance, should appear through the
variational analysis.
The effective masses for $H_g$ and $H_u$ are shown as the lower plot
in \fig{fig:g1_mass}; there is a clear separation from the mass of the
ground-state nucleon, with the negative-parity state being lower in
mass, as observed experimentally.

\begin{figure}[htb]
\psfig{width=7cm,file=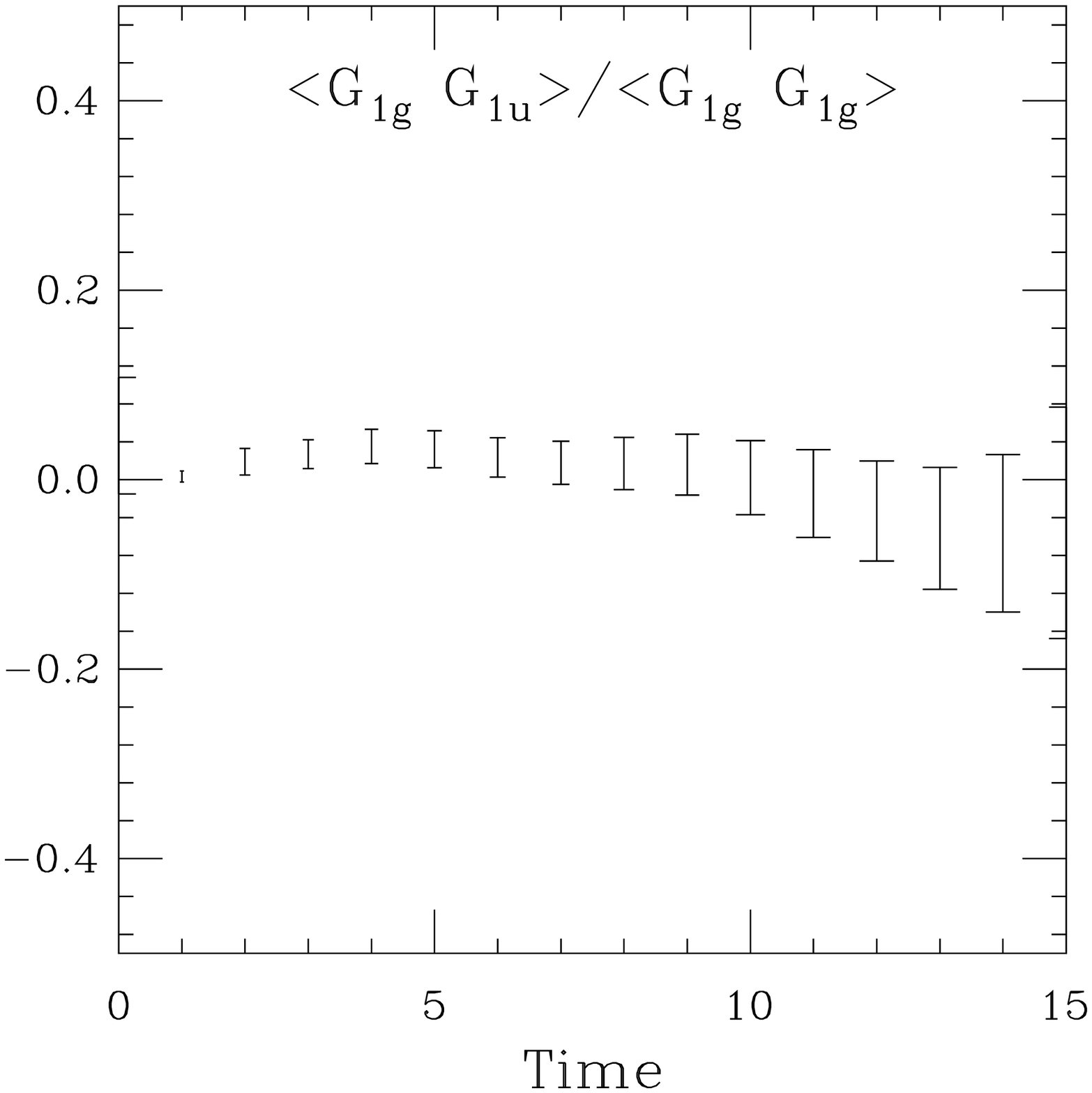}
\psfig{width=7cm,file=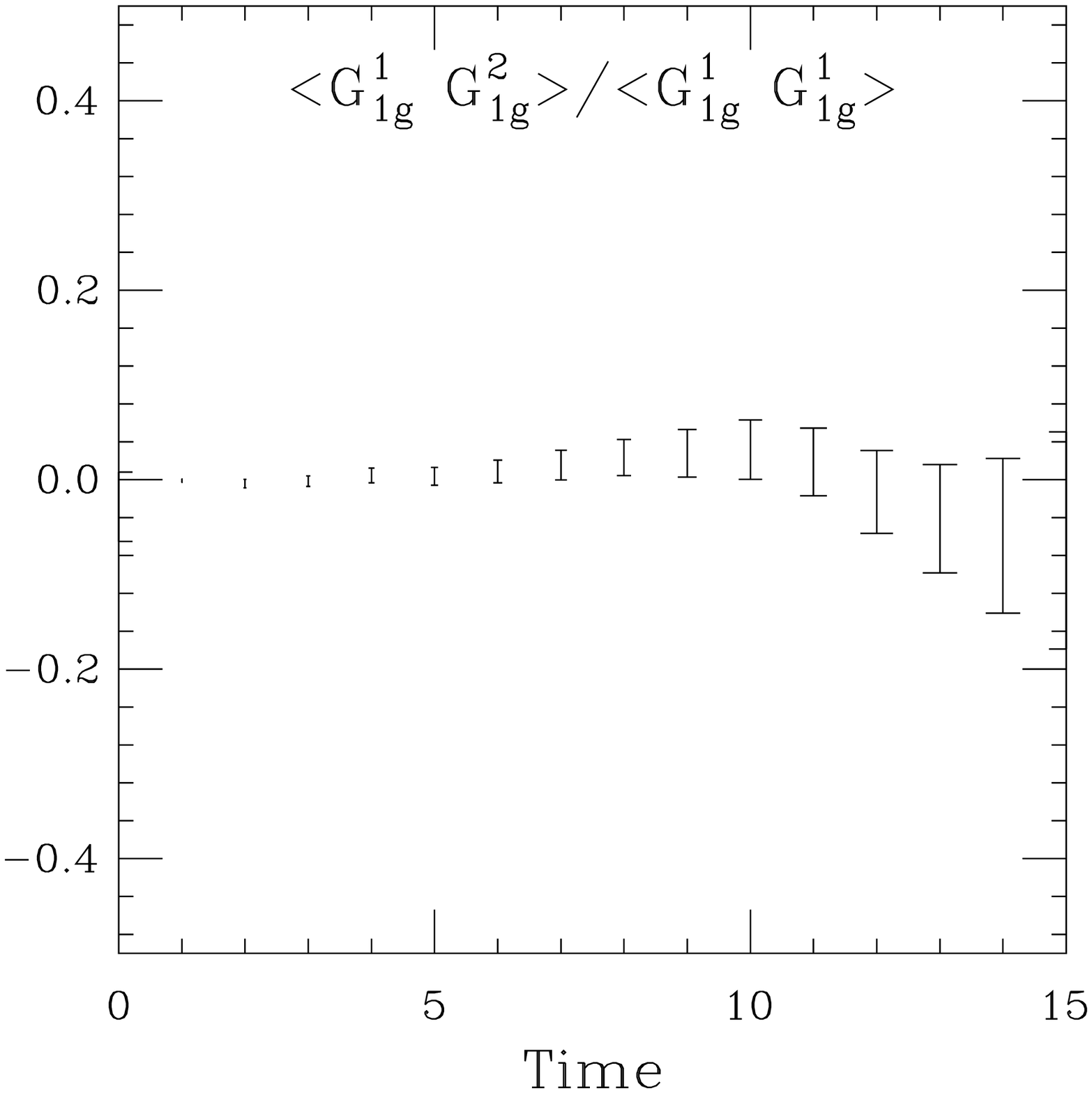}
\caption{The upper figure shows the ratio $\langle G_{1g} G_{1u}
  \rangle / \langle G_{1g} G_{1g}\rangle$,
  whilst the lower figures shows $\langle G_{1g}^{\rm row 1}
  G_{1g}^{\rm row 2} \rangle /\langle G_{1g}^{\rm row 1}
  G_{1g}^{\rm row 1} \rangle$.}
\label{fig:orthog}
\end{figure}
We illustrate the orthogonality properties in two ways in
\fig{fig:orthog}: firstly, by confirming that the cross-correlators
between different irreducible representations are indeed zero, and
secondly by verifying that those between different rows of the same
irreducible representation vanish.  The orthogonality properties are
clearly well satisfied in these representative examples, indicating
that the numerical implementation is under control.

\section{CONCLUSIONS}

We have detailed the design of operators needed to determine the excited
baryon spectrum.  The method used to construct operators transforming
irreducibly under the symmetries of the lattice has been outlined, and
the efficacy of the method illustrated.  The calculation of
``generalised'' baryon correlators followed by group-theory
projections is computationally efficient, and allows straightforward
identification of the quantum numbers of the correlators.
Furthermore, the methodology is readily extendible to
more complicated operators, such as those for pentaquarks and those
having excited glue.

This work was supported by the U.S.~National Science Foundation under
Awards PHY-0099450 and PHY-0300065, and by the U.S.~Department of
Energy under contracts DE-AC05-84ER40150 and DE-FG02-93ER-40762. We
are indebted to members of the MILC collaboration both for making
available their configurations and for many valuable insights, and to
Anna Hasenfratz and John Negele for many helpful discussions.
Computations were performed on the 128-node Pentium IV cluster at
JLab, and at ORNL, under the auspices of the US DOE's SciDAC
Initiative.


\begin{thebibliography}{10}
\bibitem{michael85} C.~Michael, Nucl.\ Phys.\ B259, 58 (1985).
\bibitem{lw90} M.~L\"{u}scher and U.~Wolff, Nucl.\ Phys.\ B339, 222 (1990).
\bibitem{APEsmear}
   M.~Albanese {\it et al.,} Phys.\ Lett.\ B {\bf 192}, 163 (1987).
\bibitem{johnson}
  R.C.~Johnson, Phys.\ Lett.\ B {\bf 114}, 147 (1982).
\bibitem{mandula2} 
  J.~Mandula and E.~Shpiz, Nucl.\ Phys.\ B{\bf 232}, 180 (1984).
\bibitem{morningstar03} R. Edwards \textit{et al} (LHP Collaboration),
hep-lat/0309079, to appear in Proceedings of Lattice 2003.
\bibitem{brommel03} Dirk Br\"{o}mmel \textit{et al.}, hep-ph/0307073.
\bibitem{milc99} C.~Bernard \textit{et al.} (MILC Collaboration),
  Phys.~Rev.~D64, 054506 (2001).
\bibitem{gray2002} A.~Gray \textit{et al,}, Nucl.\ Phys.\ (Proc.\
  Suppl.) 119, 592 (2003).
\bibitem{hyp} A.~Hasenfratz and F.~Knechtli, Phys.\ Rev.\ D64, 034504 (2001).
\bibitem{lhpc_milc} LHP Collaboration, in preparation.
\end{thebibliography}
\end{document}